\documentclass{elsart}
\usepackage{graphicx}

\begin{document}

\begin{frontmatter}
\title{Transition from Baryonic to Mesonic\\ Freeze-Out}

\author[UCT-CERN]{J.~Cleymans},
\author[TUD]{H.~Oeschler},
\author[UW]{K.~Redlich},
\author[UCT-CERN]{S.~Wheaton}
\address[UCT-CERN]{UCT-CERN Research Centre and Department  of  Physics, \\
University of Cape Town, Rondebosch 7701, South Africa}
\address[TUD]{Institut f\"ur Kernphysik,
Technische Universit\"at Darmstadt, D-64289~Darmstadt, Germany}
\address[UW]{Institute of Theoretical Physics, University of Wroc\l aw,
 Pl-45204 Wroc\l aw, Poland }
\date{\today}

\begin{abstract}
The recently discovered sharp peak in the $K^+/\pi^+$ ratio
in relativistic heavy-ion collisions is
discussed in the framework of the statistical model. In this model
a rapid change is expected as the hadronic gas undergoes a
transition from a baryon-dominated  to a meson-dominated gas. The
transition occurs at a temperature $T$ = 140 MeV and baryon
chemical potential $\mu_B$ = 410 MeV corresponding to an incident
energy of $\sqrt{s_{NN}}$ = 8.2 GeV. The maximum in the
$\Lambda/\pi$ ratio is well reproduced by the statistical model,
but the change in the $K^+/\pi^+$ ratio is much less pronounced
than the one observed by the NA49 collaboration. The
calculated smooth increase of the $K^-/\pi^-$ ratio and the shape
of the $\Xi^-/\pi^+$ and $\Omega^-/\pi^+$ ratios
exhibiting maxima at different incident energies is
consistent with the presently available experimental data. We 
conclude that the measured particle ratios with $20-30\%$ deviations
agree with a hadronic freeze-out scenario. These deviations seem
to occur just in the transition from baryon-dominated to
meson-dominated freeze-out.
\end{abstract}
\end{frontmatter}

The NA49 Collaboration  has recently  performed a series of measurements
of Pb-Pb collisions at 20, 30, 40, 80 and 158 AGeV beam energies
\cite{Gazdzicki,NA49,Lambda-NA49}.
When these results are combined with measurements at lower beam energies
from the
AGS~\cite{pi-AGS,Ahle_1999,Ahle_2000,Lambda-AGS,Ahmad,Klay} they
reveal an unusually sharp variation with beam energy 
in the $\Lambda/\left<\pi\right>$, 
with
$\left<\pi\right>\equiv 3/2(\pi^++\pi^-)$,
 and 
$K^+/\pi^+$ ratios. Such a strong variation with
energy does not occur in pp collisions and therefore indicates a
major difference in heavy-ion collisions. This transition
 has been referred to in Ref.~\cite{Gazdzicki} as the ``horn''.
A strong variation with energy of the $\Lambda/\left<\pi\right>$
ratio has been predicted on the basis of  arguments put forward
in~\cite{Gorenstein}. It has also been
suggested recently in Ref.~\cite{Stock} that this is a signal of
the special critical point of the QCD phase
diagram~\cite{Wilczek,Fodor-Katz,Tawfik} at high baryon density.
In this
paper we   explore another, less spectacular,
 possibility for the origin of the sharp maximum, namely as being due
to the transition from a baryon-dominated to a meson-dominated
hadronic gas. 
The distinction being based on whether  the entropy of the hadronic gas
is dominated by baryons or by mesons.
For this purpose we study various quantities
along the freeze-out curve~\cite{1gev} as a function of
$\sqrt{s_{NN}}$.

\begin{center}
\begin{figure}
\includegraphics[width=13.5cm]{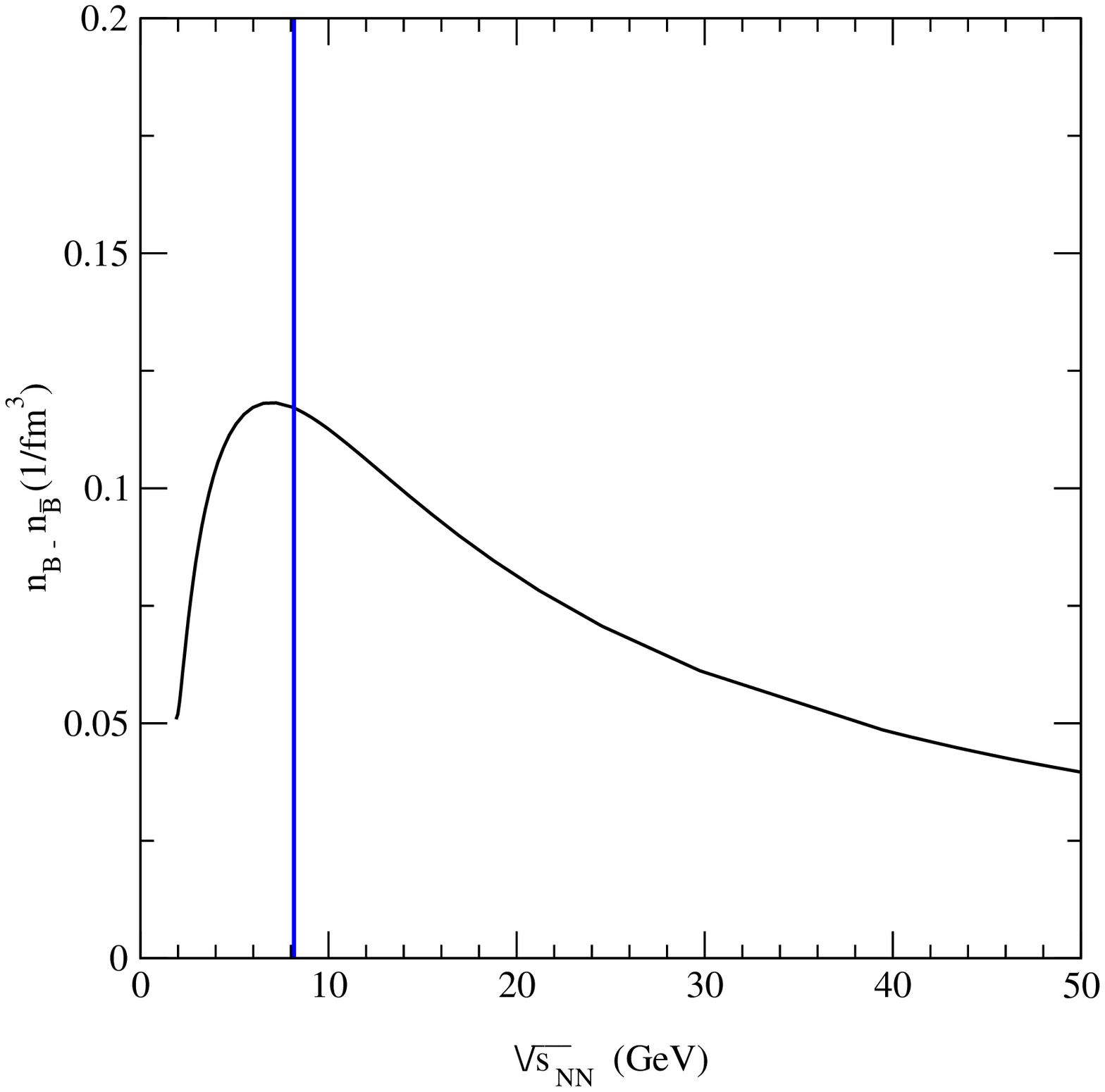}
\caption{The net baryon density as a function of $\sqrt{s_{NN}}$
calculated along the chemical freeze-out curve~\cite{1gev}
corresponding to $E/N$ = 1 GeV. } \label{netbaryon}
\end{figure}
\end{center}

In the statistical model a steep rise at low energies and a
subsequent flattening off leading to a mild maximum in the
$K^+/\pi^+$ ratio, was predicted many years
ago~\cite{cor,Becattini,max}. The sharpness of the observed 
peak therefore comes as a surprise. On the other hand, a sharp
peak in the $\Lambda/\left<\pi\right>$ ratio was predicted by the
statistical model~\cite{max} and is in good agreement with the
data. While the statistical model cannot explain the sharpness of
the peak in the $K^+/\pi^+$ ratio, there are nevertheless several
phenomena, giving rise to the rapid change, which warrant a closer
look at the model. In Fig.~\ref{netbaryon} we show the net baryon
density calculated along the chemical freeze-out curve~\cite{1gev}
as a function of $\sqrt{s_{NN}}$. This curve shows a clear maximum
with the net baryon density decreasing rapidly towards higher
energies. 

\begin{figure}[t]
\vspace*{-2.8cm}
\includegraphics[width=13.5cm]{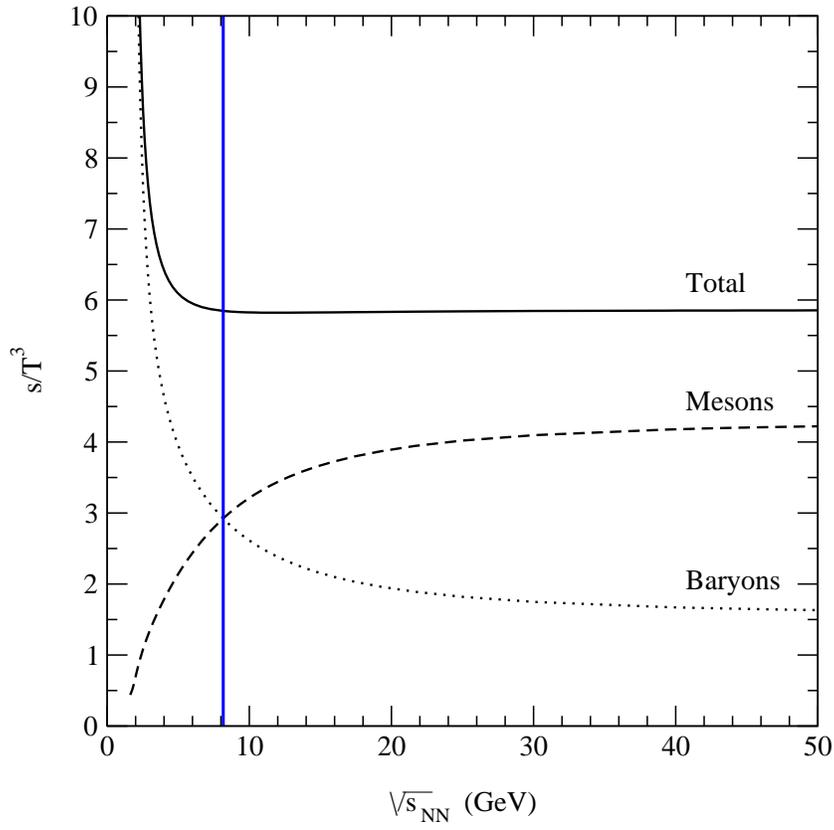}
 \caption{The entropy density normalised to $T^3$
as a function of the beam energy as calculated in the statistical model using
THERMUS~\cite{thermus}.}
\label{entropy_s}
\end{figure}

\begin{center}
\begin{figure}[ht]
\includegraphics[width=13.5cm]{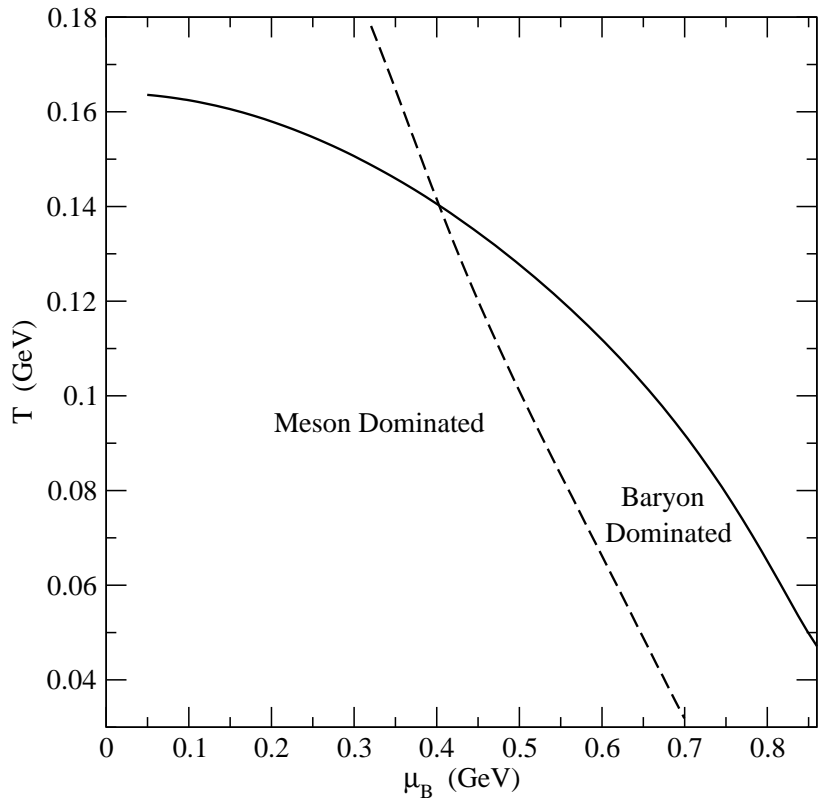}
\caption{The chemical freeze-out curve~\cite{1gev} together
with the regions where baryonic (or mesonic) contributions to the
entropy density dominate, separated by the dashed line.
Calculated in the statistical model using
THERMUS~\cite{thermus}.}
\label{entropy_pd}
\end{figure}
\end{center}

To get a better estimate of the statistical parameters in
the transition region we show in Fig.~\ref{entropy_s} the entropy
density as a function of beam energy following the freeze-out
curve given in \cite{1gev}. The separate contribution of
mesons and of baryons to the total entropy is also shown in this
figure. There is a clear  change of baryon to meson  dominance at
$\sqrt{s_{NN}}$ = 8.2 GeV. Above this value the  entropy is
carried mainly by mesonic degrees of freedom. It is remarkable that
the entropy density divided by $T^3$ is constant over the 
entire freeze-out curve, except for the  low-energy region
corresponding to the SIS energy region.
The line denoting the transition from 
a baryon-dominated to a meson-dominated hadron
gas  is shown in Fig.~\ref{entropy_pd}. This line crosses
the freeze-out curve at a temperature of $T = 140$ MeV, when the
baryon chemical potential equals $\mu_B$ = 410 MeV. The
corresponding invariant energy is $\sqrt{s_{NN}}$ = 8.2 GeV.

The strong decrease in the net baryon density seen in
Fig.~\ref{netbaryon} is due to the fact that low energies are
characterized by a very low multiplicity of mesons and,
correspondingly, a very large baryon-to-meson ratio. As a
consequence, the baryon chemical potential is also very large.
 As the beam energy is increased, meson production increases
and the baryon chemical potential decreases. The
number of strange baryons produced in heavy-ion collisions at
different collision energies will follow the net baryon density
since a large baryon chemical potential will also enhance the
number of hyperons.

\begin{center}
\begin{figure}
\includegraphics[width=13.5cm] {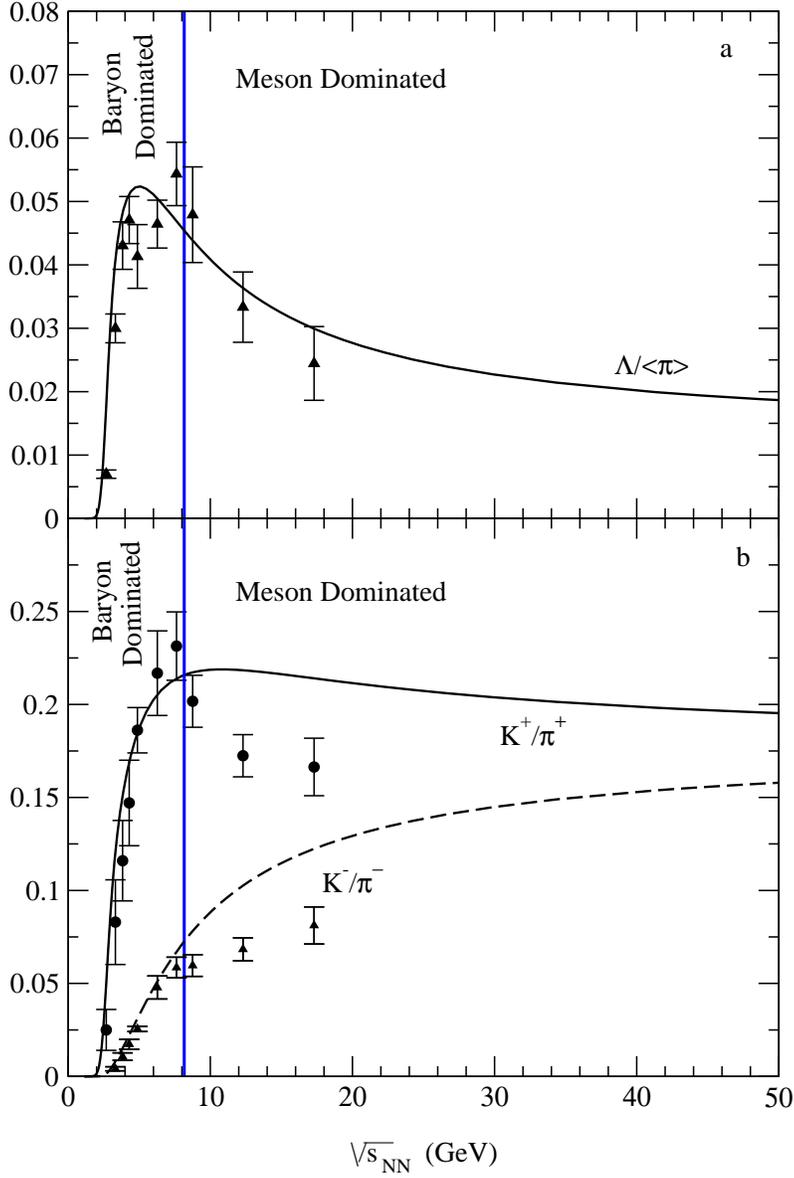}
\caption{a) The $\Lambda/\left<\pi\right>$ ratio as a function of beam energy.
b) The $K^+/\pi^+$ and $K^-/\pi^-$  ratios
as a function of energy.
The solid and dashed  lines are  the predictions of the statistical model
calculated using THERMUS~\cite{thermus}.
The data points are from Ref.~\cite{Gazdzicki,NA49,Lambda-NA49,pi-AGS,Lambda-AGS,Klay}.
}
\label{CanKpLa}
\end{figure}
\end{center}

The corresponding $K^+/\pi^+$ ratio is shown in
Fig.~\ref{CanKpLa}. As is well-known~\cite{review,manninen1}, the
statistical model description leads to a mild maximum in this
ratio which does not reproduce the so-called ``horn'' observed by
the NA49 collaboration~\cite{Gazdzicki}. 
 The observed
deviations at the highest SPS energy have been interpreted as a
lack of full chemical equilibrium in the strangeness sector,
leading to a strangeness suppression factor, $\gamma_s$, deviating
from its equilibrium value by about thirty percent. Detailed fits
using the statistical model in the region of the ``horn'' show
rapid variations in $\gamma_s$ \cite{manninen1} which do not lend 
themselves to any interpretation. There is no corresponding peak in 
the $K^-/\pi^-$ ratio because the production of $K^-$ is not tied 
to that of baryons. As the relative number of baryons
decreases with increasing energy, there is no corresponding decrease in the
number of $K^-$ as is the case with $K^+$ as these must be balanced by strange baryons.

\begin{center}
\begin{figure}[ht]
\includegraphics[width=13.5cm] {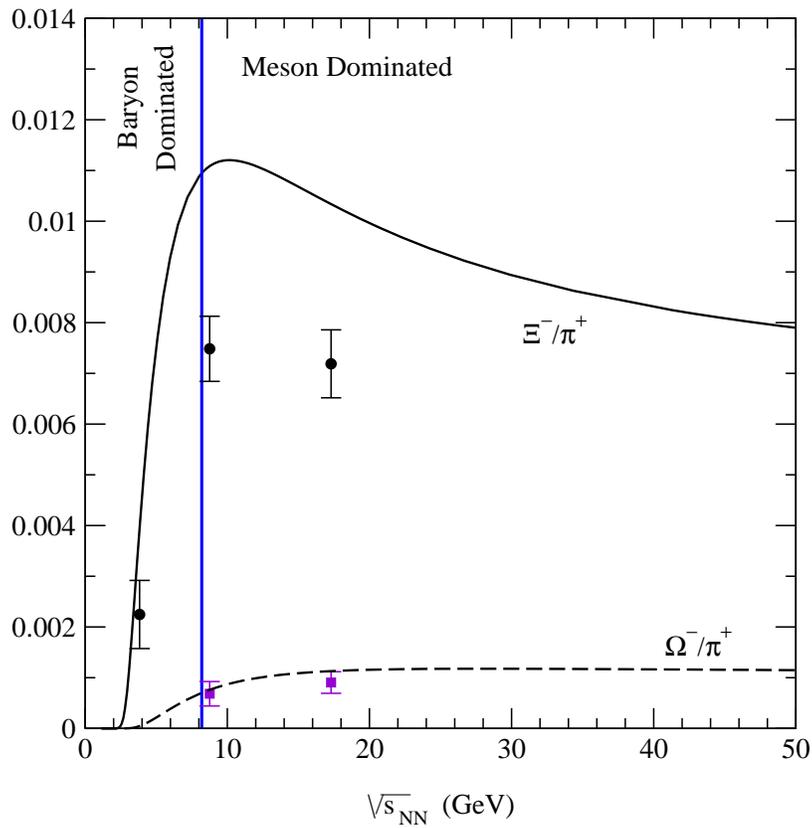}
\caption{The $\Xi^-/\pi^+$ (full line) and $\Omega^-/\pi^+$ (dashed line)
ratios as a function of beam energy calculated using THERMUS~\cite{thermus}.
The  data points are from Ref.~\cite{Xi-NA49,Xi-AGS,Omega}. The square points correspond to the
$(\Omega + \bar\Omega)/\pi^+$ ratio while the round points are for the 
$\Xi^-/\pi^+$ ratio.
}
\label{CanXiOmega}
\end{figure}
\end{center}

It is worth noting that the maxima in the
 ratios for multi-strange baryons occur at ever
higher beam energies. This can be seen clearly in
Fig.~\ref{CanXiOmega} for the $\Xi^-/\pi^+$ ratio which peaks at a
higher value of the beam energy. The ratio
$\Omega^-/\pi^+$  also shows
a (very weak) maximum, as can be seen in Fig.~\ref{CanXiOmega}. 
 The higher the strangeness content of the baryon, the higher
in energy is the maximum. This behavior is due to a combination
of the facts that strangeness has to be balanced, the baryon
chemical potential decreases rapidly with energy and the multi-strange baryons
have successively higher thresholds.
The values are listed in Table 1. 

\begin{table}
\begin{center}
\caption{ Maxima in Particle Ratios as Predicted by the Statistical Model. }
\vspace{0.5cm}
\begin{tabular}{|l|c|c|}
\hline
Ratio                       &Maximum at            & Maximum   \\
                            &$\sqrt{s_{NN}}$ (GeV)  & Value  \\
\hline                                                         
                            &                       &           \\
$\Lambda/\left<\pi\right>$  & 5.1                   & 0.052     \\
$\Xi^-/\pi^+$               & 10.2                  & 0.011     \\
$K^+/\pi^+$                 & 10.8                  & 0.22      \\
$\Omega^-/\pi^+$            & 27.0                  & 0.0012     \\
\hline
\end{tabular}
\end{center}
\end{table}
%

It is to be expected that if these maxima do not all occur at the
same temperature, i.e. at the same beam energy, then the case for
a phase transition is not very strong. The
observed behavior seems to be governed by properties of the hadron
gas. More detailed experimental studies of multi-strange hadrons
will allow the verification or disproval of the trends shown in this
paper.
It should be clear that the 
$\Omega^-/\pi^+$ ratio is very broad and shallow and it  will be difficult 
to find a maximum experimentally.

 In conclusion, while  the statistical model cannot explain
the sharpness of the peak in the $K^+/\pi^+$ ratio, its position
corresponds  precisely to a transition from a baryon-dominated to
a meson-dominated hadronic gas. This transition occurs at a
temperature $T = $ 140 MeV, a baryon chemical potential $\mu_B = $
410 MeV and an energy $\sqrt{s_{NN}} = $ 8.2 GeV. In the
statistical model this transition leads to a sharp peak in the
$\Lambda/\left<\pi\right>$ ratio, and to
 moderate peaks in the $K^+/\pi^+$, $\Xi^-/\pi^+$ and
$\Omega^-/\pi^+$ ratios. Furthermore, these peaks are at 
different energies in the statistical model. The statistical model
predicts that the maxima in the $\Lambda/\left<\pi\right>$,
$\Xi^-/\pi^+$ and $\Omega^-/\pi^+$ occur at increasing
beam energies.

If the change in properties 
of the above excitation functions were associated with
a genuine deconfinement phase transition one would expect these
changes to  occur at the same beam energy.  It is clear that more data
are needed to clarify the precise nature of the sharp variation
observed by the NA49 collaboration.
\section*{ Acknowledgments}

 We thank C. Blume for his help with the NA49 data.
We acknowledge the support of the German Bundesministerium f\"ur Bildung und
Forschung (BMBF), the Polish State Committee for Scientific
Research (KBN) grant 2P03 (06925), the
National Research Foundation (NRF, Pretoria) and the URC of the
University of Cape Town.

%
%
%
\end{document}